# Single Photon Source Using Laser Pulses and Two-Photon Absorption

B.C. Jacobs, T.B. Pittman, and J.D. Franson
*Johns Hopkins University, Applied Physics Laboratory, Laurel, MD 20723 USA*

We have previously shown that two-photon absorption (TPA) and the quantum Zeno effect can be used to make deterministic quantum logic devices from an otherwise linear optical system. Here we show that this type of quantum Zeno gate can be used with additional two-photon absorbing media and weak laser pulses to make a heralded single photon source. A source of this kind is expected to have a number of practical advantages that make it well suited for large scale quantum information processing applications.

Although two-photon absorption (TPA) is a nonlinear optical process, it is not typically considered a fundamental resource for optical quantum information processing (QIP). We have previously shown that TPA and the quantum Zeno effect can be used to make deterministic quantum logic devices (Zeno gates) from an otherwise linear optical system [1]. In a Zeno gate, TPA is used to suppress the failure events that would normally occur in a linear optics device [2-4] when multiple photons exit the device in the same optical mode. Here we show that additional two-photon absorbing media can be used in a more conventional manner along with a Zeno gate to convert weak laser pulses into heralded single photon pulses. Because recent theoretical results indicate that single photon losses can be much less than the rate of TPA [5], realistic devices of this kind could become critical components for future optical QIP systems.

There have been many demonstrations of single photon sources over the past few years using a variety of physical systems, including parametric down conversion (PDC) [6], quantum dots [7], and single molecules [8]. Two metrics commonly used to categorize these sources are the heralding efficiency, which is the probability that the output contains a single photon given a trigger signal from the source, and the production efficiency, which is the probability that the source will produce a single photon on any given attempt. Although PDC sources have demonstrated heralding efficiencies approaching 85% [6], the conversion rate of the PDC process currently limits the overall production rate to much less than 1%. This would mean that a very large number of these types of sources would have to be combined with very low loss switches in order to make a deterministic single photon source.

Unheralded sources, such as quantum dots, have demonstrated relatively high production rates ~20% [7], but their potential use in large QIP systems may be limited by the lack of a heralding signal. There have recently been several proposals to make heralding devices for these types of sources using linear [9] and nonlinear [10] optical techniques, including TPA; however, the maximum heralding efficiency using these techniques and assuming ideal nonlinearities, perfect detectors, and neglecting single photon loss, is under 85% [10]. The source we present here can be viewed as an unheralded source with an ideal production efficiency of 50% followed by a heralding circuit (Zeno gate) with an ideal heralding efficiency of 100%. The potential performance of this type of source assuming commercially available detectors and a range of TPA material characteristics is presented below.

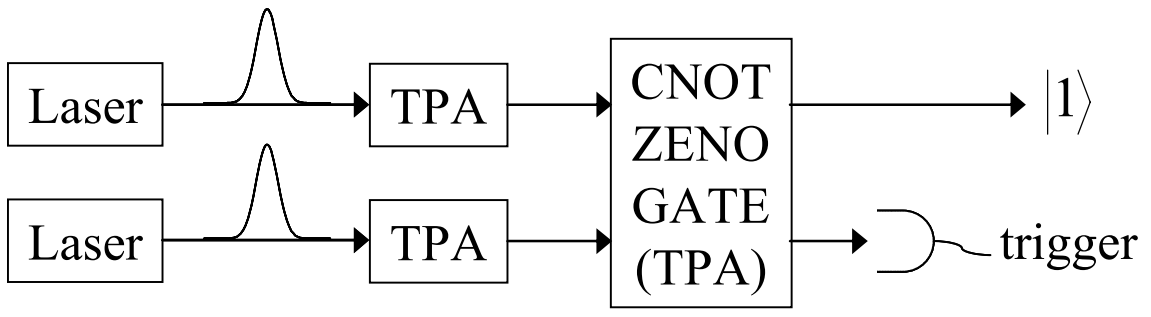

Fig. 1. Heralded single-photon source using TPA in two different ways. Prior to the Zeno gate, strong TPA is used to modify (filter) the photon number distribution of each laser pulse by absorbing photons in pairs. Inside the gate, strong TPA and the quantum Zeno effect are used to implement a CNOT gate [1]. A photo detection event signals that one photon was present in each input and that the remaining output contains a single photon.

The basic operation of the proposed source is illustrated in Fig. 1. First, strong TPA is used to convert two weak laser pulses into mixed states containing roughly equal probabilities of 0 or 1 photon with an arbitrarily small multi-photon probability. Then a quantum CNOT gate based on the Zeno effect is used to perform a quantum non-demolition (QND) measurement on one of the input modes. A detection event in one output mode indicates both the successful operation of the Zeno gate and the presence of one photon in each input mode, resulting in a heralded single-photon output. The performance of both stages of the source depend on the TPA rate $R_2$ and the single photon loss rate $R_1$. Although there are practical reasons why these characteristics might be different inside the Zeno gate, here we will assume the same characteristics throughout.

The use of optical nonlinearities to transform the Poisson number distribution of a laser pulse has previously been studied in a number of different systems [11-12]. Here we simply estimate the effect of TPA and single photon loss on a weak coherent state (WCS) $|\alpha\rangle$ by performing a density matrix calculation using a truncated representation for the initial state. Because the initial mean photon number μ was relatively low (<5), typically less than 20 states were required to keep the truncation effects negligible. Since the main goal of the initial TPA medium, or filtering cell, is to reduce the probability of more than one photon surviving, we assumed that the length L of these cells was relatively large, and that the overall TPA rate/cell $\Gamma_2=R_2L/c$ was fixed at $\Gamma_2=15$; $c$ is the speed of light. Under these conditions, and assuming a random laser phase, it is easily seen that the WCS is transformed into a mixed state given by

$$|\alpha\rangle \xrightarrow{R_1,R_2} \rho_i = P_0|0\rangle\langle 0| + P_1|1\rangle\langle 1| + \rho_{n>1}\,. \quad (1)$$

Here $P_0$ and $P_1$ are the probabilities that the filtered output contains 0 or 1 photons respectively, and $\rho_{n>1}$ represents the remaining multi-photon terms, which for $\Gamma_2=15$ occur with probability $\mathrm{Tr}[\rho_{n>1}] < 10^{-6}$.

The vacuum and single photon probabilities using this fixed TPA cell are shown in the plots of Fig. 2. When the single photon loss is low (Fig. 2a), $P_0$ and $P_1$ approach ½ for $\mu \gtrsim 3.5$ because the initial state has roughly an equal probability of having an even or an odd number of photons; under ideal TPA the even terms would collapse to vacuum and the odd terms $\rightarrow |1\rangle$. As the single photon loss increases (Fig. 2b) the distribution is skewed toward vacuum, as can be expected.

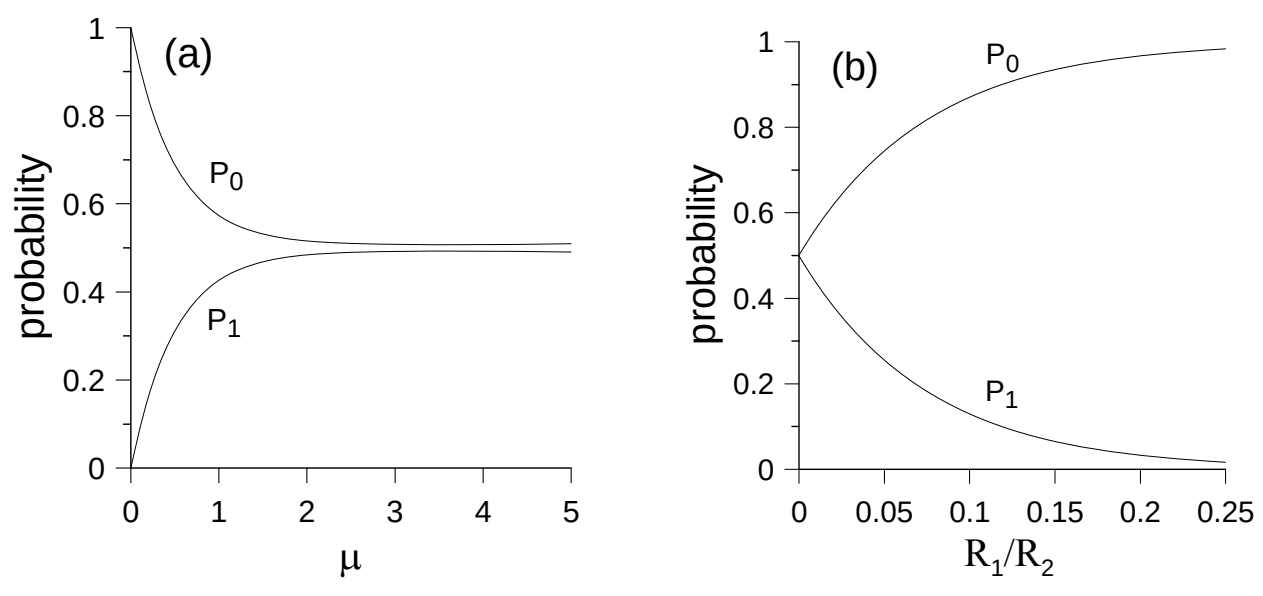

Fig. 2. Probability of a single laser pulse containing exactly zero ($P_0$) or one ($P_1$) photon after propagating through a long TPA medium as a function of (a) the initial mean photon number μ, and (b) the relative single photon loss $R_1/R_2$. In (a) $R_1/R_2 = 10^{-3}$, and in (b) $\mu = 3.5$.

After the multi-photon components of the laser pulses have essentially been removed, a heralded single-photon pulse from the upper TPA cell can be identified by performing a QND measurement using the Zeno gate circuit shown in Fig. 3. The circuit is based on a SWAP′ gate [1], whose intended function is to interchange the values of the two inputs while applying a 180 degree phase shift if a photon was present in each input. Reversing the two outputs will then produce a conventional nonlinear sign gate, which in turn will produce a CNOT operation if it is placed between two Hadamards. The presence of a photon from the upper source in the control channel $C_1$ will thus flip the target bit (when present) to produce a count in detector Dz, thereby heralding the photon in the output channel c.

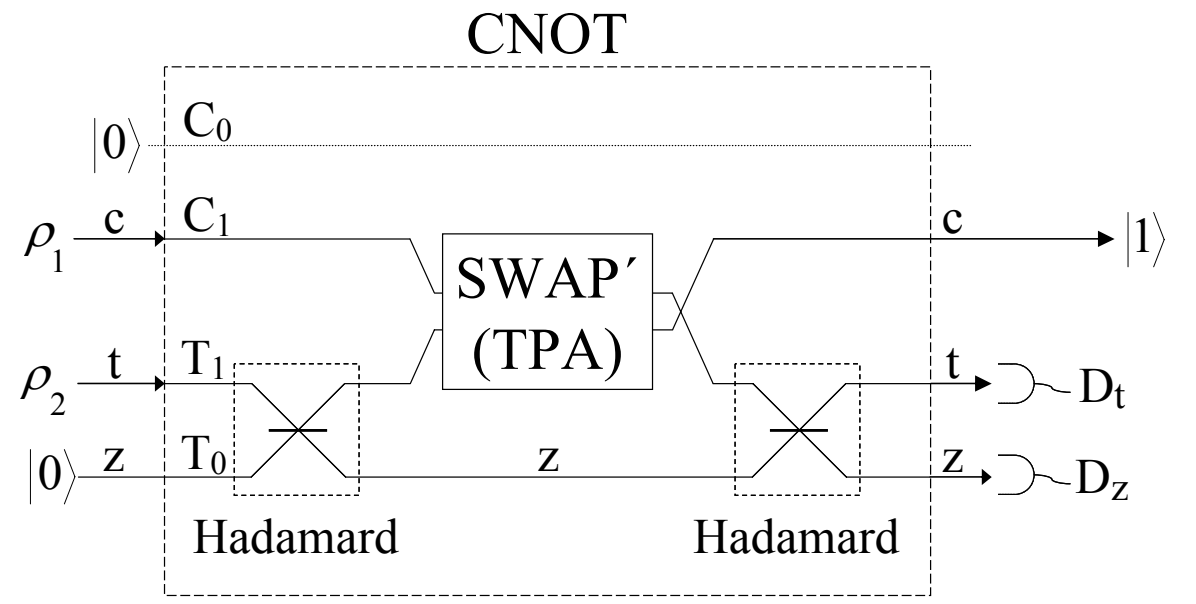

Fig. 3. Heralding circuit using a Zeno gate (SWAP′) to implement a nonlinear phase flip. The circuit is equivalent to a dual-rail encoded CNOT with mixed control and target inputs, followed by measurements of both target output modes. A detection event by detector Dz in the target logical 0 output mode z heralds the presence of a single photon in output mode c.

In order to facilitate the analysis, the circuit is presented in the context of a dual-rail encoded CNOT gate. The filtered laser pulses are input in modes c and t, which also correspond to the control ($C_1$) and target ($T_1$) logical "1" inputs respectively. The logical "0" inputs ($C_0$ and $T_0$) are left empty.

Because the input states are assumed to rarely contain more than one photon, we will only consider states that contain a maximum of two photons distributed among the optical modes c, t, and z. On input, mode z, which also corresponds to the CNOT logical $T_0$ mode, is empty, and it will be a detection event in this output channel that heralds the presence of a single photon in output mode c.

The two 50/50 beam splitters identified as Hadamard gates in Fig. 3 are assumed to transform the creation operators $\hat{z}^\dagger$ and $\hat{t}^\dagger$ according to

$$\hat{z}^\dagger \rightarrow \frac{1}{\sqrt{2}}\left(\hat{z}^\dagger + \hat{t}^\dagger\right)$$
$$\hat{t}^\dagger \rightarrow \frac{1}{\sqrt{2}}\left(\hat{z}^\dagger - \hat{t}^\dagger\right) \quad (2)$$

It can be shown that the operation of these beam splitters in the basis formed by the single photon states ($|001\rangle$, $|010\rangle$, and $|100\rangle$) is given by

$$U_1 = \begin{pmatrix} -r & 0 & r \\ 0 & 1 & 0 \\ r & 0 & r \end{pmatrix}. \quad (3)$$

Here $r = 1/\sqrt{2}$ and a ket $|zct\rangle$ indicates the number of photons in each of the corresponding modes z, c, and t. Similarly, the matrix representation of Eq. (2) in the basis formed by the two-photon states ($|011\rangle$, $|002\rangle$, $|020\rangle$, $|101\rangle$, $|110\rangle$, and $|002\rangle$ ) is

$$U_2 = \begin{pmatrix} -r & 0 & 0 & 0 & r & 0 \\ 0 & r^2 & 0 & -r & 0 & r^2 \\ 0 & 0 & 1 & 0 & 0 & 0 \\ 0 & -r & 0 & 0 & 0 & r \\ r & 0 & 0 & 0 & r & 0 \\ 0 & r^2 & 0 & r & 0 & r^2 \end{pmatrix}. \quad (4)$$

The phase conventions in Eq. (2) were chosen so that $U_1$ and $U_2$ correspond to Hadamard transformations on the target modes z ($T_0$) and t ($T_1$) for states with a single photon in these modes.

We have previously shown that TPA and the quantum Zeno effect can be used to implement the SWAP′ operation if we assume that modes c and t are weakly coupled by the interaction Hamiltonian

$$H' = \varepsilon(\hat{c}^\dagger\hat{t} + \hat{t}^\dagger\hat{c}). \quad (5)$$

Here $\varepsilon$ is the coupling strength and the interaction could correspond to evanescent coupling between the cores of two optical fibers, for example. The matrix representation of this interaction for the single and two photon states respectively is

$$\hat{H}'_1 = \varepsilon\begin{pmatrix} 0 & 1 & 0 \\ 1 & 0 & 0 \\ 0 & 0 & 0 \end{pmatrix}, \quad \hat{H}'_2 = \varepsilon\begin{pmatrix} 0 & \sqrt{2} & \sqrt{2} & 0 & 0 & 0 \\ \sqrt{2} & 0 & 0 & 0 & 0 & 0 \\ \sqrt{2} & 0 & 0 & 0 & 0 & 0 \\ 0 & 0 & 0 & 0 & 1 & 0 \\ 0 & 0 & 0 & 1 & 0 & 0 \\ 0 & 0 & 0 & 0 & 0 & 0 \end{pmatrix}.$$

If the interaction length is chosen to correspond to a time $\Delta t = \pi\hbar/2\varepsilon$, integration of Schrödinger's equation shows that this system (with no TPA) simply implements a SWAP operation on modes c and t; we have also included a $\pi/2$ phase shift on both modes to simplify the discussion. Thus, with no TPA the entire circuit in Fig. 3 simply implements the Identity, and the trigger detector Dz would never fire.

In order to calculate the actual operation of the Zeno gate we need to include the effect TPA has on the states $|020\rangle$ and $|002\rangle$ inside the gate. For this we performed a density matrix calculation of the 9-state system using the block diagonal interaction Hamiltonian

$$\hat{H}' = \begin{pmatrix} \hat{H}'_1 & 0 \\ 0 & \hat{H}'_2 \end{pmatrix}, \quad (6)$$

along with the standard commutation relation

$$\dot{\hat{\rho}} = \frac{1}{i\hbar}\left[\hat{H}', \hat{\rho}\right], \quad (7)$$

to calculate the dynamical evolution of the density matrix $\hat{\rho}$. TPA was included by assuming that the diagonal matrix elements $\hat{\rho}_{dd}$ corresponding to the two two-photon states of interest decay at a rate $R_2$ into an unspecified continuum of levels, i.e. $\dot{\hat{\rho}}_{dd} = -R_2\hat{\rho}_{dd}$. Similar decay terms to off diagonal elements involving these states were also added according to the methods in Carmichael [13]. Additionally, the effects of single photon loss ($R_1$) were included using similar techniques.

Because the required interaction time $\Delta t$ inside the Zeno gate can be varied by controlling the coupling strength, it is convenient to characterize the overall TPA strength of the Zeno gate by $\Gamma_2 = \Delta t R_2$. In the absence of single photon loss, i.e. $R_1 = 0$, the results of numerically integrating the coupled differential equations above were identical to our previous results [1]. This indicates that strong TPA in this system prevents two photons in separate modes ($|110\rangle$) from entering the same optical mode. Furthermore, the results indicate that this state experiences a phase flip, i.e., $|110\rangle \xrightarrow{\text{SWAP}'} -|110\rangle$.

In order to evaluate the potential usefulness of this approach we performed a density matrix calculation of the entire source. For simplicity we assume that input states consisting of more than two photons always produce errors, and that the detectors can be completely described by their quantum efficiency $\eta$ and probability of producing a dark count $P_d$. Additionally, we assume the presence of a detector in the $T_1$ output mode to suppress incorrect heralding events due to failures in the Zeno gate. The overall single photon production rate $P_s$ and error rate assuming ideal ($\eta=1$, $P_d=0$) and realistic ($\eta=0.75$, $P_d=10^{-5}$) detectors is presented in Fig. 4, where the single-photon loss is shown as a fraction of $R_2$.

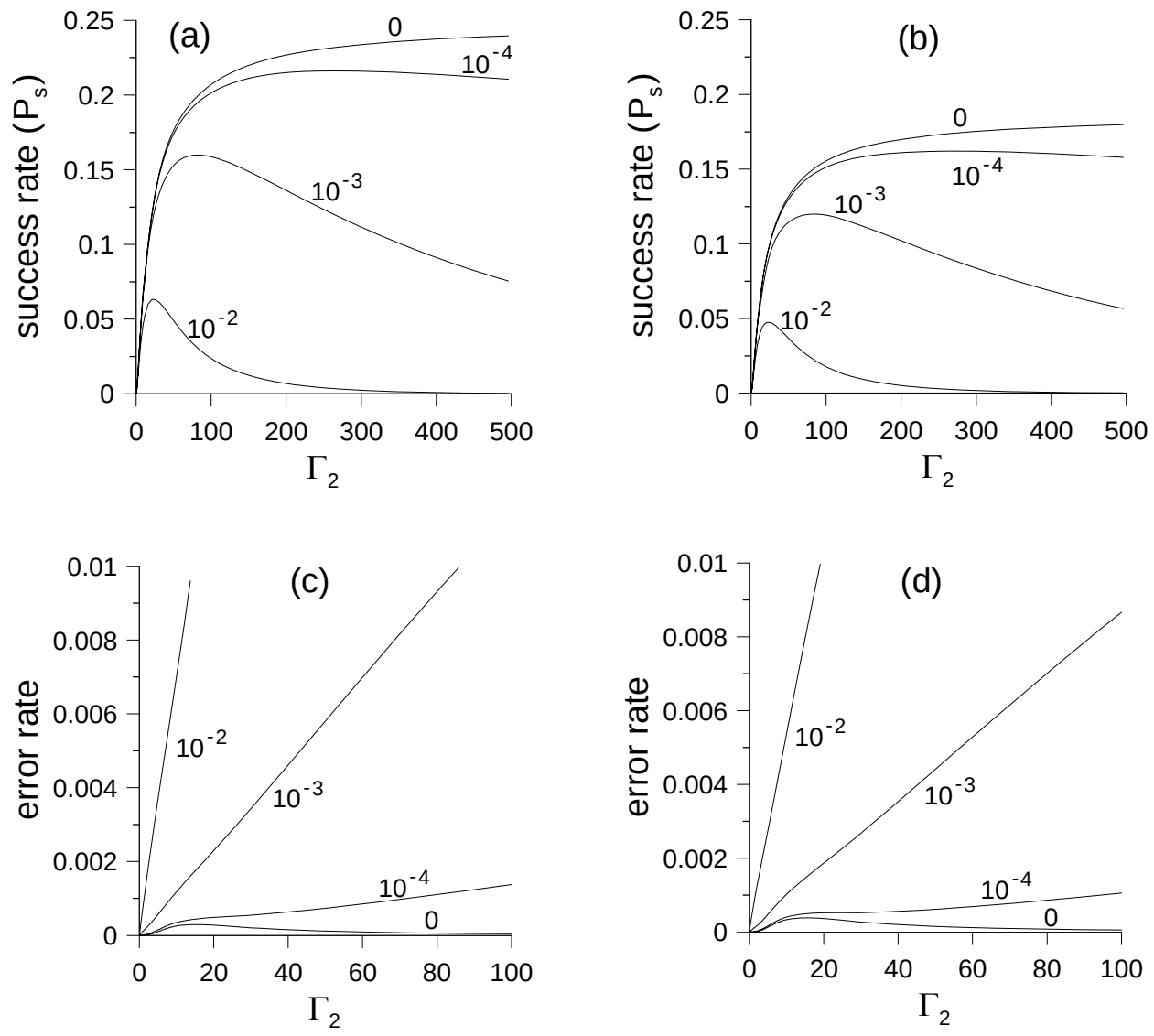

Fig. 4. Single photon production efficiency (top) and error rate (bottom) as a function of the Zeno gate TPA strength ($\Gamma_2$) given four relative strengths {0, $10^{-2}$, $10^{-3}$, and $10^{-4}$} of single photon loss. Ideal results assuming perfect detectors are shown on the left and commercially available detector characteristics are assumed for the plots on the right.

For clarity, $P_s$ is the probability that the source produces and heralds a perfect single photon output on any given shot, and the error rate $P_e$ is simply the probability of a false trigger. Additionally, the heralding efficiency *H*, which is the conditional probability of success, is given by $H = P_s/(P_s + P_e)$. For comparison with other types of sources, the fidelity $F \equiv \sqrt{\langle 1|\rho|1\rangle} = \sqrt{H}$ of this source is shown in Fig. 5. All of these results indicate that the performance is not substantially degraded by the detector assumptions; however, they clearly suggest the need for strong TPA with low single-photon loss.

Remarkably, it can be seen that the heralding circuit continues to function with relatively low error rates (at reduced efficiency) as $\Gamma_2 \rightarrow 0$. This is because when only one photon is present the operation of the circuit does not depend on TPA, and there is no false triggering- aside from detector dark counts. Additionally, as $\Gamma_2$ vanishes so does the nonlinear sign flip for the two-photon state, and the circuit simply implements the Identity, again limiting the false trigger events to dark counts.

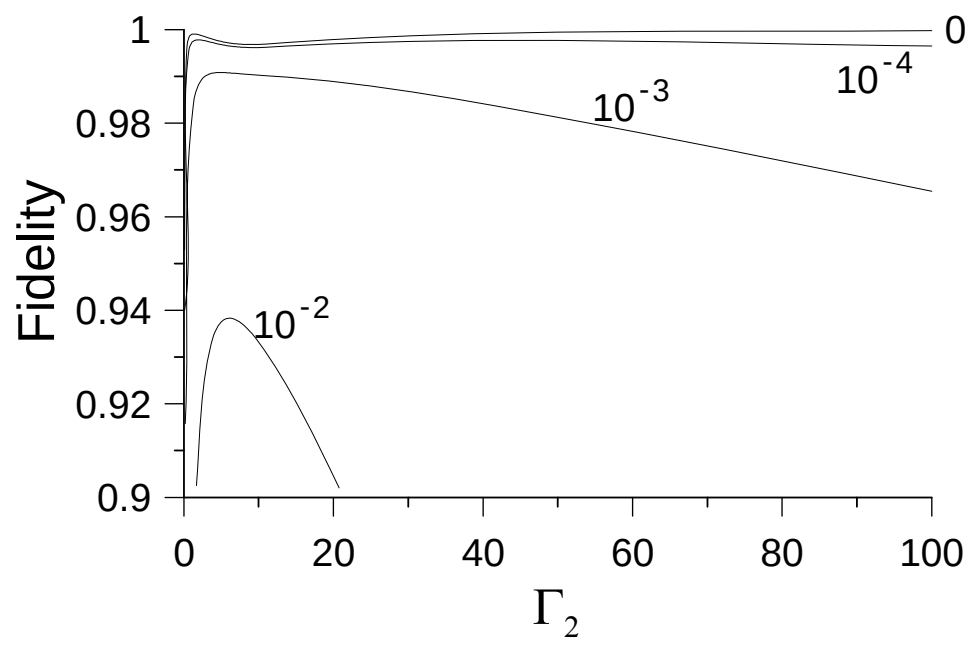

Fig. 5. Overall source fidelity as a function of the Zeno gate TPA strength ($\Gamma_2$) assuming commercially available detectors. The Zeno gate and initial TPA filtering medium are assumed to have the same relative single photon loss, as listed, but the filtering medium length is chosen to provide a fixed level ($\Gamma_2$=15) of TPA.

In summary, we have shown that TPA can be used to robustly convert laser pulses into heralded single-photon pulses. From Ref. 5, single-photon losses of $10^{-3}$ should be achievable, in which case single-photon outputs with fidelity above 98% could be efficiently produced using this method.

This work was supported by DTO, ARO, and IR&D funding.